# Discovering Knowledge from Multi-modal Lecture Recordings

Rajkumar. K [1], Christian Gütl [2,3],

[1] Department of Computer Science, Bishop Heber College, Tiruchirappalli 620017, India
[2] Institute for Information Systems and Computer Media (IICM), TU-Graz, Austria
[3] Infodelio Information Systems and GÜTL IT Research & Consulting, Austria
[1] rajkumarkannan@{yahoo.co.in, demfoundation.org}, [2] cguetl@iicm.edu

**Abstract**

*Educational media mining is the process of converting raw media data from educational systems to useful information that can be used to design learning systems, answer research questions and allow personalized learning experiences. Knowledge discovery encompasses a wide range of techniques ranging from database queries to more recent developments in machine learning and language technology. Educational media mining techniques are now being used in IT Services research worldwide. Multi-modal Lecture Recordings is one of the important types of educational media and this paper explores the research challenges for mining lecture recordings for the efficient personalized learning experiences.*

*Keywords: Educational Media Mining; Lecture Recordings, Multimodal Information System, Personalized Learning; Online Course Ware; Skills and Competences;*

## 1. Introduction

The rapid evolution of communication technologies, advanced data networks and the semi-automation of multimedia information processing have greatly impacted learning experiences. Multimedia resources include not only entertainment media such as movie, cartoon, advertisement, computer games and so on, but also educational media such as available in digital library, museum, digital archives, e-learning systems etc. Lecture recording, like courses and presentations, is one of the important educational media types. Further information of multi-media resources and their application in e-Learning can be found in literature such as [6].

Within the last years universities have increasingly started to record lectures and offer these recordings students for their learning procedures. Some universities offer even complete series of recordings of hundreds of courses available for public access such as MIT OCW [8] and Berkeley courses on YouTube [2].

Most of the aforementioned lecture recordings are indeed multi-modal in nature. Most of them include audio and video stream(s) as well as an information channel for synchronized presentation slides. Additionally, information about the interaction with the presenter PC (e.g. next slide, mouse movements and like) and interaction with other media can add further useful information. Moreover, the linkage with lecture notes, course outline and background information are further information which should be taken into account. Also teacher-student interaction and student inputs are sources of information to be kept and analyzed. Further information about multimodal information systems can be found for example in [4].

Such media recordings provide a wide range of useful information acquisition within the learning process (see for example [9]) but also for analyzing purposes (see for example [1]). Therefore Educational Media Mining is a vital area of research because it facilitates

- Detecting affect and disengagement
- Detecting attempts to circumvent learning called "gaming the system"
- Guiding student learning efforts
- Developing or refining student models
- Measuring the effect of individual interventions
- Improving teaching support
- Predicting student performance and behavior

However, we need to develop standard data formats and ontology, so that researchers can more easily share data and conduct meta-analysis across systems, and use pre-existing systems for storing, managing, accessing, searching and visualizing media recordings and linked metadata. Furthermore, we need to determine which data mining techniques are most appropriate for the specific features of educational data, and how these techniques can be used on a large scale.

In this paper, we explore the foremost challenges for educational media mining in particular lecture recordings for the efficient personalized learning experiences.

## 2. Challenges for Mining Lecture Recordings

Recently, there is a growing interest in capturing the live lecture presentations for subsequent distribution to students at distant locations, besides organizing the course content as a collection of web pages [5] [7]. Hence, incorporating knowledge and discovery in distance learning environments becomes important and unavoidable. Therefore, the foremost challenges for the mining of lecture recordings are to enable the effective
- Creation
- Integration
- Exploration
- Analysis
- Presentation of knowledge from web based content, so that the users can enjoy the facilities.

For this, we need an intelligent system that will capable of organizing lecture recordings automatically into technically exploitable knowledge source and providing users with information access environment to utilize the multimedia content.

## 3. Phases in Mining Multi-modal Lecture Recordings

Personalized learning is a demanding requirement for any learning systems, see for example [5] and [3]. For a learner, it customizes the learning material based on their learning preferences in the context of culture, content and so on. For an efficient personalized learning experience, one should know:
- What sort of information should be extracted from lecture media.
- What knowledge can be derived from this information.
- How to put the information into a form that is suitable for students' learning experience.

By focusing to lecture recordings, students want either to view lectures they have missed, view again problematic parts or to retrieve just segments which are in the context the current learning task. Therefore, a modern system must be able deal with these different situations.

The above requirements lead us into the different phases of the mining process. The mining process can be broadly categorized as shown below:
- Annotation
- Knowledge discovery
- E-learning environment
- Knowledge repository

### Annotation

Instructional repositories archive enormous amount of multimedia data as lecture recordings with the support of advanced digital technology. Since lecture media is highly unstructured, an efficient access cannot be possible without the proper indexing scheme. Lack of such indexing scheme results inefficient search and browse of the lecture notes and recordings. Since manual indexing is tedious, automatic indexing scheme is highly important. Techniques for transforming lecture recordings into meaningful structured information can be done by
- Classifying lecture video content automatically
- Recognizing key events in the delivery of the lecture
- Summarization.

### Knowledge Discovery

Educational media mining is fundamentally different from other general data mining methods. It should integrate the existing data mining technologies with the educational theories to improve the performance of our intelligent system. During educational media mining, one should consider the following questions, which will guide the discovery process. They are,
- What techniques are especially useful for data mining
- How can data mining improve trainer support
- How can data mining build better student models
- How can data mining dynamically alter instruction more effectively for personalized learning

### E-learning Environment

The intelligent system should possess a good learning environment for personalized content learning upon request. It should have state-of-the-art human-computer interaction methods and visualization techniques to interface with the system. The user interface can contain a PC in intranet or internet set up. It can also integrate mobile devices such as touch pad, audio and visual display. Besides, hardware based dedicated chip with the necessary sensors and other devices would be noteworthy environment for the intelligent system.

### Knowledge Repositories

In order to make use of the discovered structures from the media recordings, the management of these structures is an essential technological goal. The management includes addition, editing and deleting of ontological elements and relations between those elements.

## 3. Conclusion

Educational media mining is emerged as a new way of discovering facts from the diverse educational media objects more recently. In this paper, particular attention has been paid to lecture recordings and their application for learning purposes. Based on that, challenges and the different phases in the mining process have been discusses. Currently, we have started to build on a prototype dealing with lecture recordings.